\documentclass{iau} 
\usepackage{graphicx}

\title[GRBs, AGN jets and dark matter] 
{Polarimetric studies of GRBs, AGN jets, and axion dark matter}

\author[K. Toma]   
{Kenji Toma$^{1,2}$
}

\affiliation{$^1$Frontier Research Institute for Interdisciplinary Sciences, Tohoku University, Sendai 980-8578, Japan\\
  $^2$Astronomical Institute, Graduate School of Science, Tohoku University, Sendai 980-8578, Japan \\ email: {\tt toma@astr.tohoku.ac.jp} \\[\affilskip]
}

\pubyear{2021}
\volume{360}  
\jname{Astronomical Polarimetry 2020: New Era of Multiwavelength \\ Polarimetry}
\editors{H.Shinnaga, B-G Andersson, A.M.Magalh\~{a}es  \& E.Falgarone, eds.}
\begin{document}

\maketitle

\begin{abstract}
  {
Relativistic jets are collimated outflows with speeds close to light speed, which are associated with gamma-ray bursts (GRBs), active galactic nuclei (AGNs), and so on. This article mainly overviews recent developments of polarimetric studies of GRBs and their afterglows in the gamma-ray and optical wavebands as well as the first detections of their radio polarization. Polarimetric observations and theoretical modelings can address the emission mechanism, magnetic field structure, and energetics of GRB jets and related collisionless plasma physics. Some of the discussed key physics are common with AGN jets. Furthermore, we mention that polarimetry of AGN jets and protoplanetary disks may be a novel approach to search for ultra-light axion dark matter.    
  }
\end{abstract}

\firstsection 
\section{Introduction and summary}
\label{sec:introduction}

\begin{figure}[t]
  \begin{center}
          \includegraphics[width=5.5in]{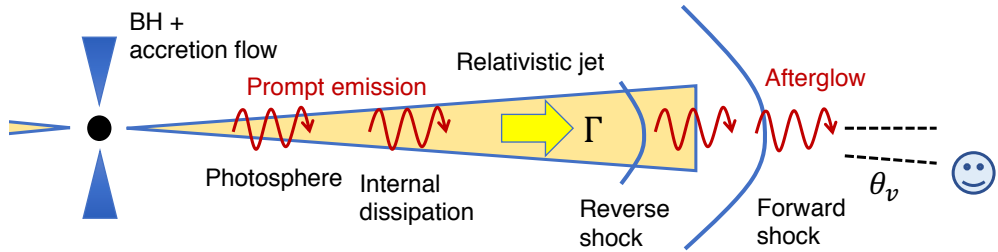}
 \caption{Schematic picture of physical mechanism of GRBs, where $\Gamma$ and $\theta_v$ represent the bulk Lorentz factor of jet and the viewing angle (angle between the jet axis and the sight line), respectively.}
   \label{fig:grb}
\end{center}
\end{figure}

Gamma-ray bursts (GRBs) are the brightest transients in the gamma-ray sky, and their origins have been an active research target, related with multi-messenger astronomy (e.g., \cite[Murase et al. 2006]{murase06}; \cite[Inoue, Granot et al. 2013]{inoue_s13}; \cite[Abbott et al. 2017]{abbott17}; \cite[Kimura et al. 2017]{kimura17}) and observational cosmology (for a review, \cite[Toma et al. 2016]{toma16b}). The current observational data indicate a promising scenario that core collapse of a massive star or coalescence of a compact star binary results in the system of a highly spinning black hole (BH) surrounded by an accretion disk and produces a relativistic jet. The jet emits the gamma-rays, and then the interaction of the jet with the circumburst medium generates the reverse shock (RS) which propagates within the jet and the forward shock (FS) which propagates in the circumburst medium, both of which emit the long-lived multi-band afterglows (see Figure~\ref{fig:grb} and \cite[M\'{e}sz\'{a}ros 2002]{meszaros02}; \cite[Piran 2004]{piran04}; \cite[Kumar \& Zhang 2015]{kumar15} for reviews).

However, this scenario has some major problems: One is whether the jets are driven by thermal neutrino energy released from the accretion disk (\cite[Eichler et al. 1989]{eichler89}; \cite[Zalamea \& Beloborodov 2011]{zalamea11}) or by electromagnetic energy extracted from the rotating central BH (\cite[Blandford \& Znajek 1977]{blandford77}; \cite[Komissarov 2004]{komissarov04}; \cite[Toma \& Takahara 2016]{toma16}; \cite[Kimura et al. 2021]{kumura21}). Prompt gamma-ray emission mechanism is still also a mystery, which should have essential information on the particle and energy contents of jets (e.g., \cite[Porth \& Komissarov 2015]{porth15}; \cite[Bromberg \& Tchekhovskoy 2016]{bromberg16}). The afterglows are known as synchrotron emission of non-thermal electrons accelerated by the collisionless shocks, while the magnetic field configuration at the shocks is not understood, which is an essential ingredient in physics of non-thermal particle acceleration (for a recent review, \cite[Sironi et al. 2015]{sironi15}).

Magnetic fields play essential roles for all these problems, and then polarimetric observations are powerful for addressing them (for reviews, \cite[Lazzati 2006]{lazzati06}; \cite[Toma 2013]{toma13}; \cite[Covino \& G\"{o}tz 2016]{covino16}; \cite[Gill et al. 2021]{gill21}). The polarimetric study is now a frontier of GRB study: A number of discoveries have been achieved recently in the gamma-ray, optical, and radio wavebands, as reviewed in the following sections. We review the current progress (as well as the future projects) of observations and related theories of the FS afterglows in Section \ref{sec:forward}, the RS afterglows in Section~\ref{sec:reverse}, and the prompt gamma-ray emission in Section~\ref{sec:gamma}. We often refer to literature on active galactic nucleus (AGN) jets, since some of the key physics are common. We also mention ideas for axion dark matter search recently proposed by using polarimetric observations of photons in Section~\ref{sec:axion}.

\section{FS afterglows of GRBs}
\label{sec:forward}


\underline{{\it Theories}}. Late-phase afterglows (say, later than $\sim 10^4\;$s after the prompt emission) are usually explained by synchrotron emission of electrons accelerated by FS with strong turbulent magnetic field. For \textit{ordered} magnetic field, synchrotron emission of electrons with isotropic momentum distribution has linear polarization degree (PD) $\sim 70\%$ and polarization direction perpendicular to both the magnetic field and the line of sight (\cite[Rybicki \& Lightman 1979]{rybicki79}). The circular PD is suppressed to the order of $\sim 1/\gamma_e$, where $\gamma_e$ is Lorentz factor of the emitting electrons (\cite[Melrose 1980a]{melrose80a}; \cite[1980b]{melrose80b}). For the turbulent magnetic field, the net PDs become lower. Thus the observed PDs reflect magnetic field configuration in the visible regions.

If the magnetic field is generated by plasma instabilities such as Weibel instability, magnetic field direction should be random on plasma skin depth scales (\cite[Medvedev \& Loeb 1999]{medvedev99}; \cite[Kato 2005]{kato05}; \cite[Keshet et al. 2009]{keshet09}; \cite[Tomita et al. 2019]{tomita19}; \cite[Asano et al. 2020]{asano20}). If the strong magnetic field is generated by magnetohydrodynamic (MHD) instabilities such as Richtmyer-Meshkov instability, the field coherence length scale should be many orders of magnitude larger than the plasma scales (\cite[Sironi \& Goodman 2007]{sironi07}; \cite[Inoue, Shimoda et al. 2013]{inoue13}). Then for isotropic turbulence, the net linear PD is $\Pi_L \sim 70\%/\sqrt{N}$, where $N$ is the number of coherent patches in the relativistically-beamed visible region of angular size $\sim 1/\Gamma$, and $\Gamma$ is the bulk Lorentz factor of shocked fluid (\cite[Gruzinov \& Waxman 1999]{gruzinov99}; \cite[Jones \& O'Dell 1977]{jones77}). Therefore it vanishes for the isotropic plasma-scale turbulent magnetic field.

However, the turbulence of the downstream of shock can be anisotropic with respect to the shock normal, and then the net linear PD can be sizable even in the plasma-scale turbulent field (\cite[Sari 1999]{sari99}; \cite[Ghisellini \& Lazzati 1999]{ghisellini99}; \cite[Lazzati 2006]{lazzati06}; \cite[Shimoda \& Toma 2021]{shimoda21}). In such models, if the angular distribution of the jet energy is uniform, the net PD is zero in the early phase when $1/\Gamma \ll \theta_j$ ($\theta_j$ is the jet opening half-angle) so that the visible region is symmetric, but it becomes non-zero for the viewing angle $\theta_v \neq 0$ after $1/\Gamma$ gets large and the jet edge makes the visible region asymmetric. Its position angle (PA) flips by 90 degree when $1/\Gamma \sim \theta_j$. The PA flips do not necessarily occur for the structured jets (\cite[Rossi et al. 2004]{rossi04}) and for the shocked fluids which consist of turbulent field and ordered field (\cite[Granot \& K\"{o}nigl 2003]{granot03}).


\underline{{\it Optical and radio polarimetric observations}}. The late-phase afterglows in the optical band typically show $\Pi_L \sim 1-3\%$ (\cite[Covino et al. 2003]{covino03}). The PA flips by 90 degrees were observed in some GRBs such as GRB 121024A and GRB 091018 (\cite[Wiersema et al. 2014]{wiersema14}), while not observed in some other GRBs such as GRB 020813 (\cite[Lazzati et al. 2004]{lazzati04}).
The degeneracy of magnetic field structure, angular structure of jet, and viewing angle in the model has not allowed us to obtain a firm conclusion. Very recently, thanks to the detection of gravitational wave of the progenitor system, the afterglow of GRB 170817A was densely observed (\cite[Abbott et al. 2017]{abbott17}), and the angular structure and viewing angle of the jet is highly constrained. For such a case the anisotropy of the magnetic field can be studied in detail (\cite[Gill \& Granot 2020]{gill20}).

\begin{figure}[t]
  \begin{center}
    \begin{tabular}{c}
      \begin{minipage}{0.5\hsize}
        \begin{center}
          \includegraphics[width=2.3in]{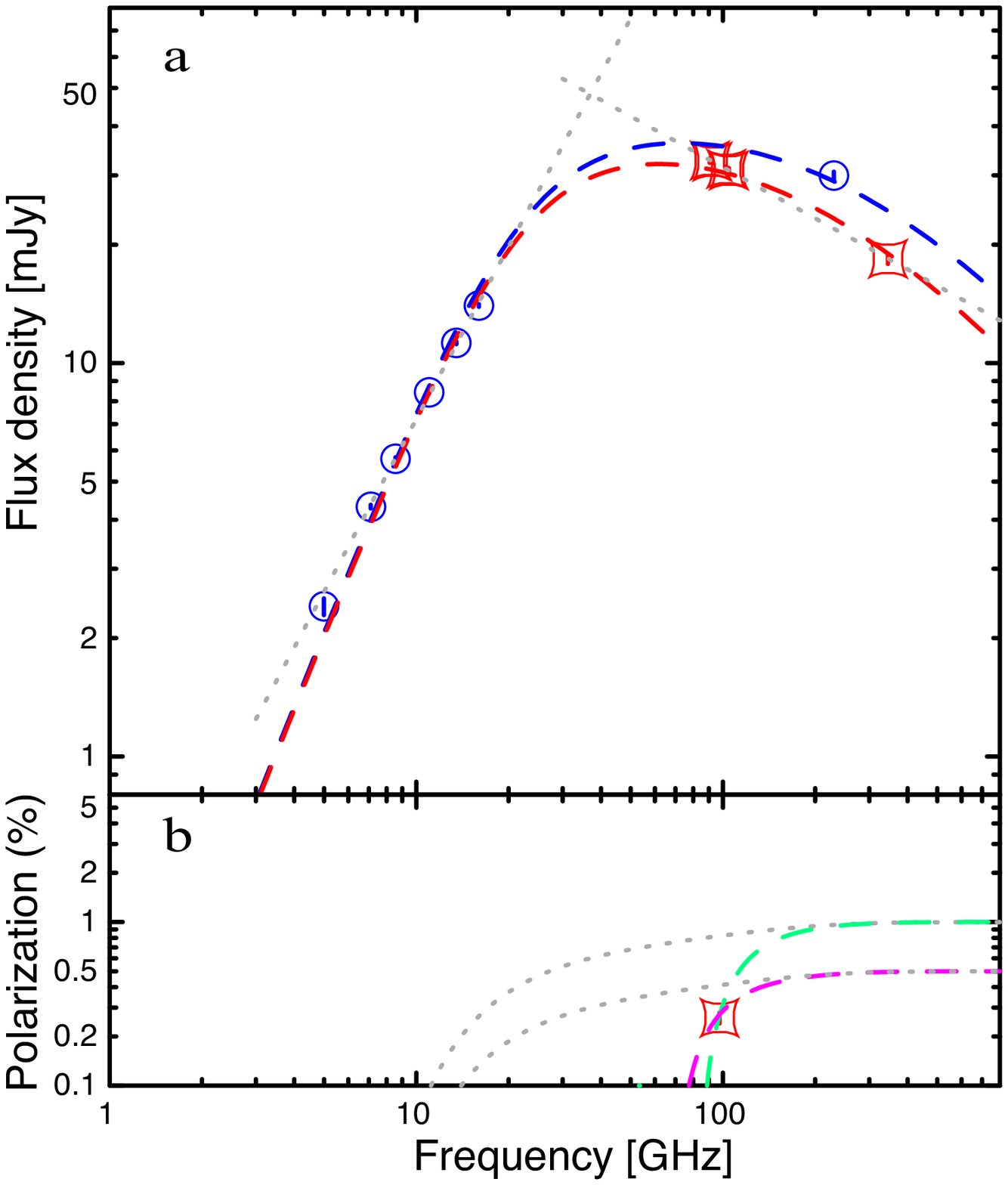}
        \end{center}
      \end{minipage}
      \begin{minipage}{0.5\hsize}
        \begin{center}
          \includegraphics[width=2.6in]{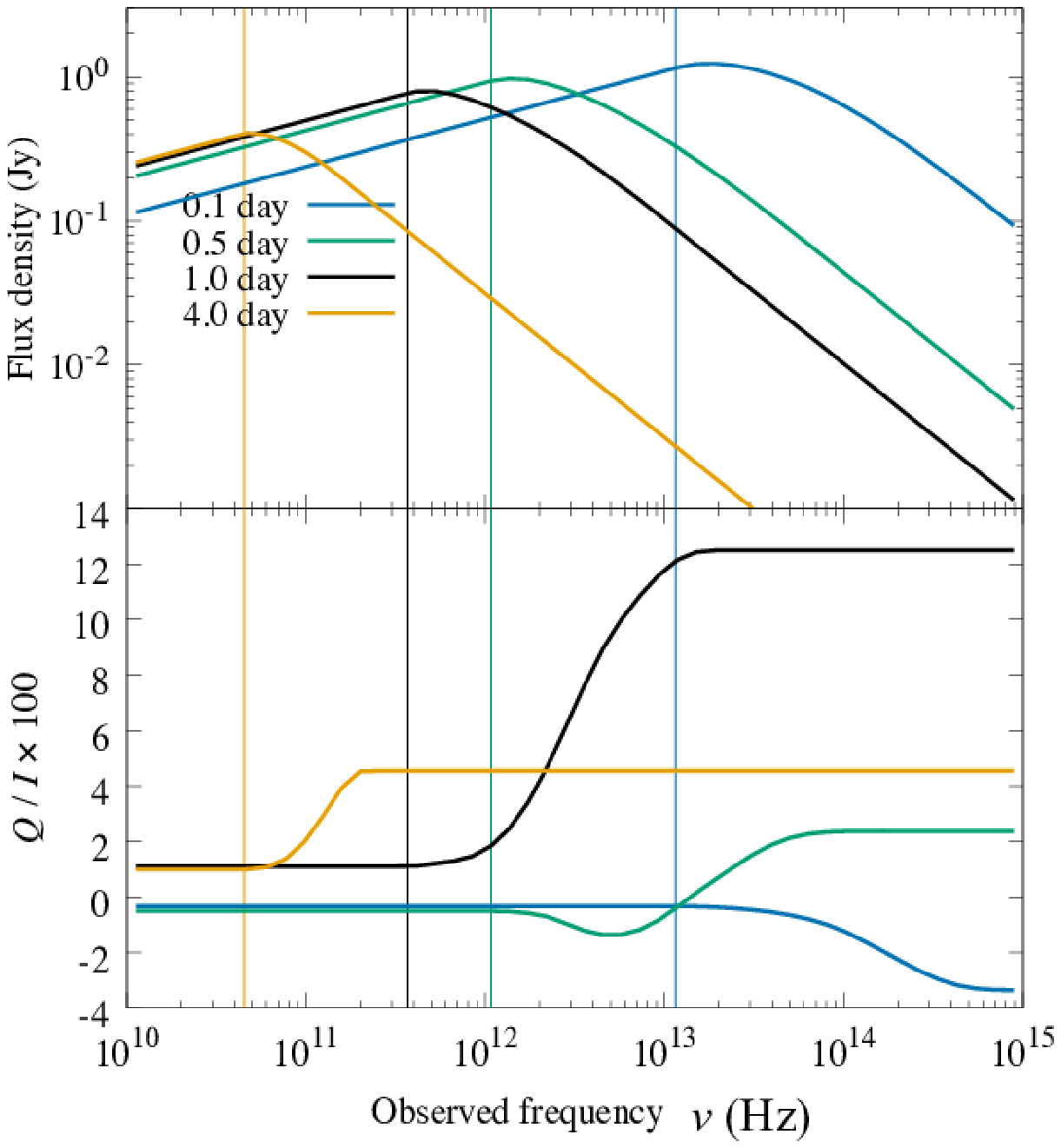}
        \end{center}
      \end{minipage}
    \end{tabular}
 \caption{{\it Left}: Intensity spectra at $4.1\;$day (blue circles) and $5.2\;$day (red squares) and PD at $5.2\;$day of GRB 171205A (\cite[Urata et al. 2019]{urata19}). {\it Right}: Intensity spectra of a GRB afterglow set at $100\;$Mpc at different days (upper panel) and their corresponding spectra of PD ($Q/I \times 100\%$) in the anisotropic plasma-scale magnetic field model, where the Stokes $U=0$ and the $\pm$ signs of $Q$ correspond to $90^\circ$ difference of PAs (bottom panel) (\cite[Shimoda \& Toma 2021]{shimoda21}).}
   \label{fig:radio}
\end{center}
\end{figure}

Kanata telescope performed the early-phase polarimetry of GRB 091208B, which appears to be FS emission and has $\Pi_L = 10.4 \pm 2.5\%$ (\cite[Uehara et al. 2012]{uehara12}). This does not favor the plasma-scale random field model explained above. It may imply the hydrodynamic-scale random field model or contribution from highly polarized RS emission (see Section~\ref{sec:reverse}).

A mysterious result was reported that one of the two GRBs with deep circular polarimetric observations with Very Large Telescope showed $\Pi_C = 0.61 \pm 0.13\%$ (\cite[Wiersema et al. 2014]{wiersema14}). This value is several orders of magnitude higher than prediction by the standard models explained above (\cite[Matsumiya \& Ioka 2003]{matsumiya03}; \cite[Sagiv et al. 2004]{sagiv04}; \cite[Toma et al. 2008]{toma08}). Dust scattering effects during the propagation is thought to be weak because the linear polarization appears typical (i.e., does not seem to be affected by dust). Therefore, this high circular PD may be intrinsic, and suggests that extremely anisotropic momentum distribution of electrons is required in the shocked fluid (\cite[Wiersema et al. 2014]{wiersema14}; \cite[Nava et al. 2016]{nava16}).

Late-phase polarimetric observations in the radio wavebands were tried for several GRBs, but no detection had been reported until recently. This is probably because the observational frequency $9\;$GHz of the previous polarimetries with VLA was below the synchrotron self-absorption frequency, for which the emission was optically thick and unpolarized (\cite[Granot \& Taylor 2005]{granot05}). Recently, the first detection of radio linear polarization has been reported for GRB 171205A by using ALMA Band 3 ($97.5\;$GHz), which was optically thin (\cite[Urata et al. 2019]{urata19}). The linear PD was $\Pi_L = 0.27 \pm 0.04\%$, which is rather low compared to typical late-phase optical PDs $\sim 1-3\%$ (no simultaneous optical polarimetric data for this GRB; see Figure~\ref{fig:radio} {\it left}). The plasma-scale random field leads to radio PD lower than optical PD because of the different intensity image shapes (see Figure~\ref{fig:radio} {\it right}), but is at odds with observed PA variation with frequencies (\cite[Shimoda \& Toma 2021]{shimoda21}; \cite[Rossi et al. 2004]{rossi04}; \cite[Granot et al. 1999]{granot99}). Instead, the low $\Pi_L$ may suggest Faraday depolarization effect within the emitting region (\cite[Urata et al. 2019]{urata19}; \cite[Melrose 1980b]{melrose80b}; \cite[Sokoloff et al. 1998]{sokoloff98}). The Faraday depolarization can be strong if a fraction $1-f$ of electrons swept by the FS are just isotropized and remain non-accelerated (\cite[Toma et al. 2008]{toma08}). If this is the case, the total GRB outflow energy is $1/f$ times larger than the usual estimate with $f=1$ (\cite[Eichler \& Waxman 2005]{eichler05}). ALMA observations of GRBs are ongoing, and further systematic studies on various GRBs would be one of the important topics of the future planned radio projects such as ngVLA and ALMA2 (e.g., \cite[Urata \& Huang 2021]{urata21}).

\section{RS afterglows of GRBs}
\label{sec:reverse}

The optical RS emission flux can overwhelm the FS flux in the early phase (say $t < 10^3\;$s; \cite[Zhang et al. 2003]{zhang03}), so that rapid observations are crucial for detecting the RS component. The RINGO2 imaging polarimeter on the Liverpool Telescope successfully detected linear polarization of many early-phase afterglows, which include seemingly RS components. They show a tendency that the RS components have higher $\Pi_L$ than the FS ones (\cite[Steele et al. 2017]{steele17}). A remarkable result is the observation of GRB 120308A, which shows high $\Pi_L$ with the maximum value $28 \pm 4\%$ with almost temporally constant PA (\cite[Mundell et al. 2013]{mundell13}). These suggest that ordered magnetic fields {\it within the jets} are not too weak to be fully distorted by the RS (\cite[Deng et al. 2017]{deng17}). A very recent radio polarimetry of the RS component of GRB 190114C with ALMA, however, indicates $\Pi_L = 0.87 \pm 0.13\%$ with gradually changing PA (\cite[Laskar et al. 2019]{laskar19}).

\section{Prompt gamma-ray emission of GRBs}
\label{sec:gamma}


\underline{{\it Theories}}. Prompt emission of GRBs exhibits large temporal variability, which is probably created within jets. Promising emission mechanisms from arguments on the intensity spectrum and radiative efficiency are photospheric quasi-thermal emission and synchrotron emission (\cite[Kumar \& Zhang 2015]{kumar15}). Polarimetric data will allow us to pin down the emission mechanism, and help understand the driving mechanism of GRB jets.

Polarization properties in the synchrotron models with MHD turbulent magnetic field (\cite[Gruzinov \& Waxman 1999]{gruzinov99}) and plasma-scale turbulent field (\cite[Granot 2003]{granot03a}; \cite[Nakar et al. 2003]{nakar03}) are similar to those of the FS afterglows (see \cite[Toma 2013]{toma13}).
Another possibility is that the ordered (probably helical) magnetic field is advected to the gamma-ray emission zone from the central engine, for which the net $\Pi_L$ is $\sim 40\%$ for on-axis viewing angles $\theta_v < \theta_j$ (\cite[Lyutikov et al. 2003]{lyutikov03}; \cite[Granot 2003]{granot03a}; \cite[Cheng et al. 2020]{cheng20}; \cite[Lan \& Dai 2020]{lan20}). 
The photospheric emission from a local region of the jet can also have linear polarization as high as $\sim 40\%$ (\cite[Beloborodov 2011]{beloborodov11}; \cite[Ito et al. 2014]{ito14}; \cite[Lundman et al. 2014]{lundman14}). The photon intensity distribution in the fluid rest frame can be anisotropic around the photosphere, and the last Thomson scattering of the photons makes them polarized. However, the PAs of the local emissions around the sight line are symmetric, the high net $\Pi_L$ is obtained only for the off-axis viewing angles, similar to the synchrotron model with plasma-scale random field (\cite[Parsotan et al. 2020]{parsotan20}; \cite[Gill et al. 2021]{gill21}). 

As expected from the above theories, Monte Carlo simulations of the observed linear polarization from uniform jets with random viewing angles show that $\Pi_L$ is clustered at $\Pi_L \sim 40\%$ for the synchrotron model with ordered magnetic field, while at $\Pi_L < 10\%$ for the synchrotron model with plasma-scale random field and the photospheric emission models (\cite[Toma et al. 2009]{toma09}; \cite[Gill et al. 2020]{gill20b}). Similar results are obtained for structured jets (\cite[Gill et al. 2020]{gill20b}).


\underline{{\it Gamma-ray and optical observations}}. The first gamma-ray polarimetric detector designed for GRBs is GAP onboard IKAROS solar power sail demonstrator. It measured the linear polarizations of three very bright bursts in the $70-300\;$keV range (\cite[Yonetoku et al. 2011]{yonetoku11}; \cite[2012]{yonetoku12}; \cite[Toma 2013]{toma13}). The polarization detection significance is $\sim 3-4\sigma$, and $\Pi_L > 30\%$ at $2\sigma$ confidence level for GRB 110301A and GRB 110721A, whose gamma-ray durations are relatively short. GRB 100826A with long duration $\sim 150\;$s shows temporal change of PA. The observed high $\Pi_L$ for bright bursts favors the synchrotron emission model with ordered field, while the synchrotron model with plasma-scale random field and the photospheric model require fine tunings. The synchrotron model with hydrodynamic-scale random field may also be viable. \cite[Inoue et al. (2011)]{inoue11} claimed that $\Pi_L \sim 70\%/\sqrt{N}$, with $N \sim 10^3$ indicated by MHD simulations of internal shocks, is not sufficient for the measured values, although this theoretical estimate of $\Pi_L$ does not take account of anisotropy of the field directions (\cite[Inoue, Shimoda et al. 2013]{inoue13}).

The IBIS detector on the INTEGRAL satellite obtained the data similar to the GAP results (\cite[G\"{o}tz et al. 2009]{gotz09}; \cite[2013]{gotz13}). More recently, the POLAR detector on Chinese Tiangong 2 Space Lab performed linear gamma-ray polarimetry of several GRBs (\cite[Zhang et al. 2019]{zhang19}; \cite[Kole et al. 2019]{kole19}). The detailed time-resolved analysis of intensity and polarization of GRB 170114A shows the spectra consistent with synchrotron emission, $\Pi_L \sim 30\%$, and PA change within a pulse (\cite[Burgess et al. 2019]{burgess19a}). The CZTI detector on AstroSat also measured a PA change in GRB 160821A (\cite[Sharma et al. 2019]{sharma19}). COSI, a balloon borne spectro-polarimeter put an upper limit $\Pi_L \lesssim 40\%$ for GRB 160530A in $0.2-2\;$MeV (\cite[Lowell et al. 2017]{lowell17}). As for recent analysis results of gamma-ray polarimetric data, readers are referred to \cite[Kole et al. (2020)]{kole20} and \cite[Chattopadhyay (2021)]{chatto21}.

In the optical band, the upper limits on prompt emission linear PD of GRB 140430A were obtained as $<30\%$ at $3\sigma$ (\cite[Kopa\v{c} et al. 2015]{kopac15}). For GRB 160625B, a hint of optical polarization detection $\Pi_L > 8\%$ was reported (\cite[Troja et al. 2017]{troja17}).

More statistical argument will be possible after the detector POLAR2 (\cite[Kole et al. 2019]{kole19}) and/or LEAP (\cite[McConnell 2017]{mcconnell17}) are launched, which will perform detailed polarization measurements of $\sim 50$ GRBs per year. Recent detailed analyses show that many GRB spectra can be fitted by synchrotron emission model (\cite[Burgess et al. 2020]{burgess19b}; \cite[Ravasio et al. 2018]{ravasio18}), while few spectra, e.g., GRB 090902B, can be fitted only by quasi-thermal model (\cite[Abdo et al. 2009]{abdo09}). It would be interesting if the bursts with synchrotron-like spectra show high $\Pi_L$ while those with thermal-like spectra show low $\Pi_L$.

\underline{{\it Implications from AGN jet polarization}}. The PA change within a pulse has been observed also in blazars in the optical band. In some cases, the PA displays smooth, monotonic rotations by as high as hundreds of degrees (e.g., \cite[Marscher et al. 2008]{marscher08}; \cite[Sasada et al. 2011]{sasada11}; \cite[Blinov et al. 2016]{blinov16}). The mechanisms of PA changes might be common between GRB prompt emission and blazar optical emission. We should note that phase effects in the combination of steady emission and burst components with slight PA changes may lead to an apparent large angle rotation (\cite[Cohen \& Savolainen 2020]{cohen20}). Recent particle-in-cell simulations show that the accelerated particles streaming through the plasmoid created by magnetic reconnection can result in a PA rotation (\cite[Zhang et al. 2018]{zhang18}; \cite[Hosking \& Sironi 2020]{hosking20}).

\begin{figure}[t]
  \begin{center}
    \begin{tabular}{c}
      \begin{minipage}{0.5\hsize}
        \begin{center}
          \includegraphics[width=2.5in]{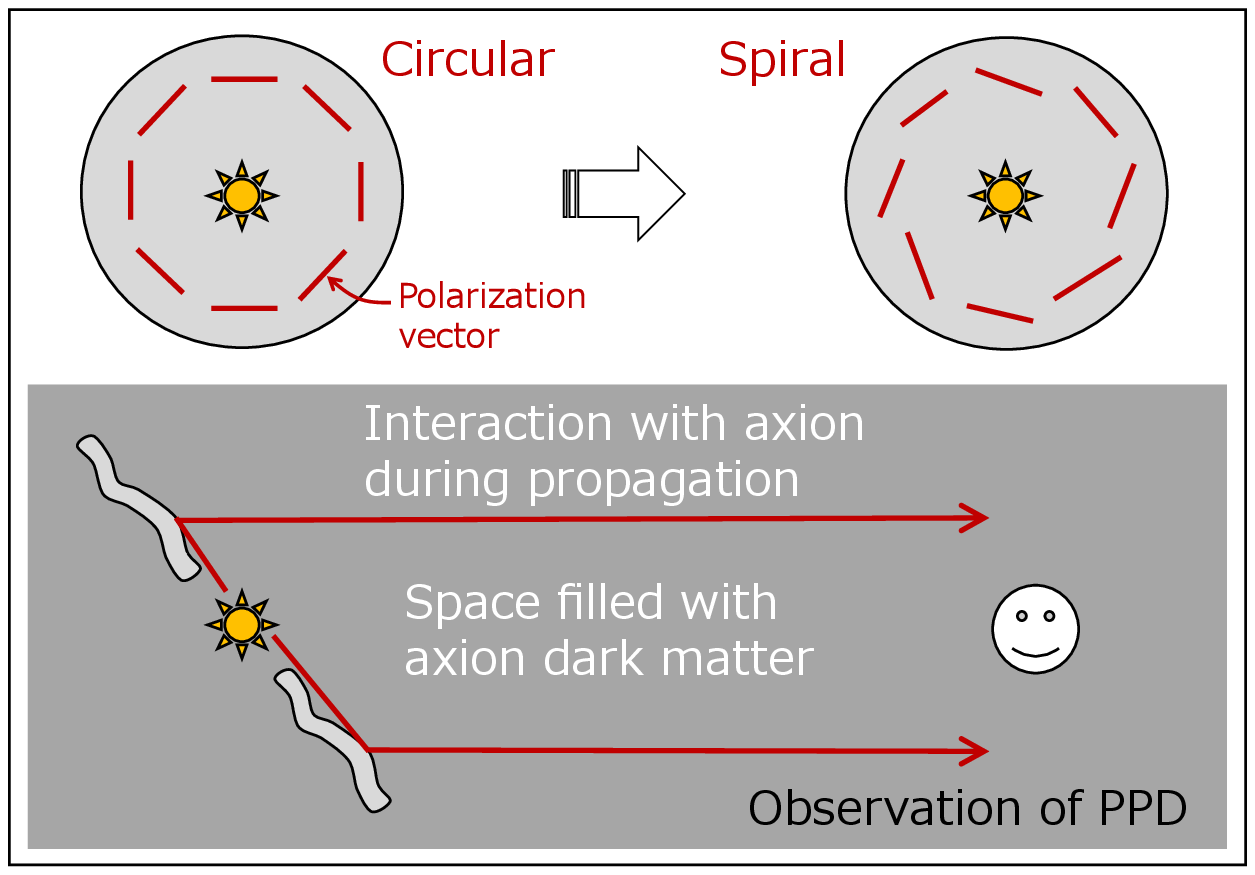}
        \end{center}
      \end{minipage}
      \begin{minipage}{0.5\hsize}
        \begin{center}
          \includegraphics[width=2.6in]{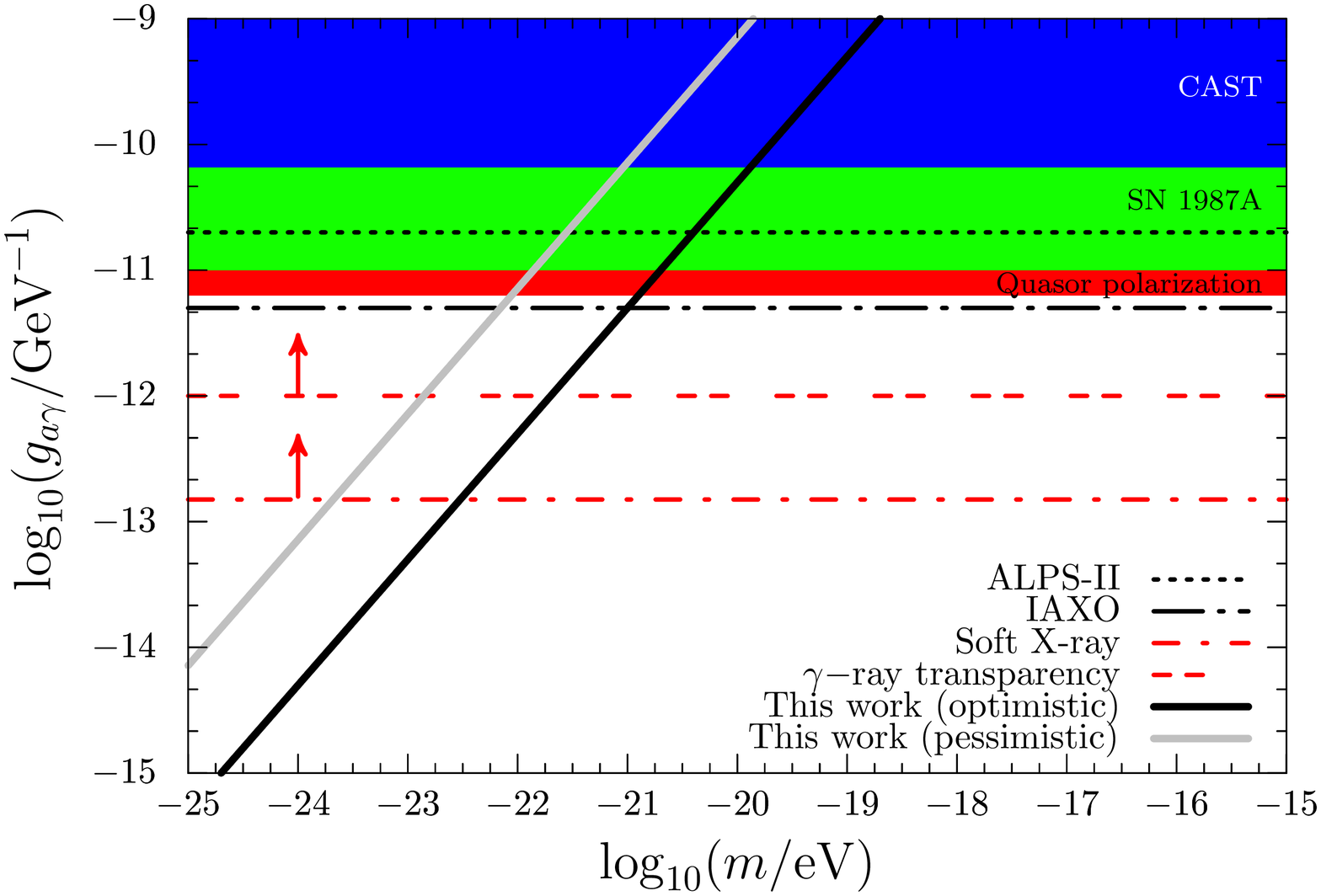}
        \end{center}
      \end{minipage}
    \end{tabular}
 \caption{{\it Left}: Schematic picture of the axion dark matter search with PPD observation. {\it Right}: Optimistic (black solid line) and pessimistic (gray solid line) upper bounds on the axion-photon coupling constant $g_{a\gamma}$ for various mass $m$ of the axion dark matter with the data of PPD AB Aur. The blue, green, and red regions are rejected by laboratory experiment and astronomical observations. The red broken lines are presumptive lower limits from X-ray and $\gamma$-ray observations (\cite[Fujita et al. 2019]{fujita19}).}
   \label{fig:axion}
\end{center}
\end{figure}

\section{Probe of fundamental physics}
\label{sec:axion}

\underline{{\it $CPT$ invariance}}. Polarimetric observations of astrophysical objects can be used for investigating fundamental physics. One example is verification of $CPT$ invariance. If $CPT$ invariance is violated as some quantum gravity theories predict, then group velocities of photons with right-handed and left-handed circular polarizations may differ slightly, which leads to birefringence and a gradual PA rotation of linear polarization. This is analogous to Faraday rotation effect on photons propagating through magnetized plasmas (\cite[Rybicki \& Lightman 1979]{ribicki79}), although the $CPT$ invariance violation effect is stronger in higher energy ranges in the opposite way to the Faraday rotation. The detections of gamma-ray linear polarization of GRBs averaged in the observed energy bands enable us to obtain strict limits on $CPT$ violation (\cite[Toma et al. 2012]{toma12}; \cite[G\"{o}tz et al. 2013]{gotz13}; \cite[2014]{gotz14}).

\underline{{\it Axion dark metter}}. Second example is axion dark matter search. Among many candidates for dark matter, ultra-light axion has recently received great attention (for a review, e.g., \cite[Irastorza \& Redondo 2018]{irastorza18}). Axion is predicted by particle physics including string theory, and can resolve the astrophysical ``core-cusp problem", which is a tension between observations and simulations of dark matter profile at galactic centers. A remarkable property of axion is the weak coupling with photons, which leads to Faraday rotation-like effect, although the PA rotation by axion does not depend on photon frequency (see Figure~\ref{fig:axion} {\it left}). \cite[Fujita et al. (2019)]{fujita19} proposed that proto-planetary disks (PPDs), whose intrinsic linear polarization pattern is known to be circular one in the near-infrared band (due to scattering of stellar photons by dust grains), are a best object for search for PA rotation by axion. They put a strong constraint on axion dark matter with polarimetric data of AB Aurigae (\cite[Hashimoto et al. 2011]{hashimoto11} and see Figure~\ref{fig:axion} {\it right}).

The rotation of PA by axion dark matter is predicted to oscillate with period $\sim 1\;$yr. Search for this oscillation has been performed for blazars (\cite[Ivanov et al. 2019]{ivanov19}), pulsars (\cite[Caputo et al. 2019]{caputo19}), and cosmic microwave background (\cite[Fedderke et al. 2019]{fedderke19}), and led to similar constraints as the studies with the PPD. It was pointed out that the axion search with PPDs have a potential for achieving the best sensitivity (\cite[Chigusa et al. 2020]{chigusa20}).

\end{document}